\documentclass[%
reprint,
superscriptaddress,
amsmath,amssymb,
aps,
]{revtex4-2}
\usepackage{graphicx}
\usepackage{dcolumn}
\usepackage{bm}
\usepackage[hidelinks]{hyperref} 
\hypersetup{
colorlinks = true,
linkcolor = blue,
anchorcolor = blue,
citecolor = blue,
urlcolor = blue
}
\usepackage{enumitem} 

\begin{document}
	
	\preprint{APS/123-QED}
	
	\title{Gravitational-wave evolution of newborn magnetars with different deformed structures}
	
	\author{Jun-Xiang Huang}
	\affiliation{Guangxi Key Laboratory for Relativistic Astrophysics, School of Physical Science and
		Technology,
		Guangxi University, Nanning 530004, China}
	
	\author{Hou-Jun L\"{u}}
	\email{lhj@gxu.edu.cn}
	\affiliation{Guangxi Key Laboratory for Relativistic Astrophysics, School of Physical Science and
		Technology,
		Guangxi University, Nanning 530004, China}
	
	\author{Jared Rice}
	\affiliation{Department of Physics, Texas State University, San Marcos, Texas 78666, USA}
	
	\author{En-Wei Liang}
	\affiliation{Guangxi Key Laboratory for Relativistic Astrophysics, School of Physical Science and
		Technology,
		Guangxi University, Nanning 530004, China}
	
	\date{\today}
	
	\begin{abstract}

Weak and continuous gravitational-wave (GW) radiation can be produced by newborn magnetars with
deformed structure and is expected to be detected by the Einstein telescope in the near
future. In this work we assume that the deformed structure of a nascent magnetar is not caused by a
single mechanism but by multiple time-varying quadrupole moments such as those present in
magnetically induced deformation, starquake-induced ellipticity, and accretion column-induced
deformation. The magnetar loses its angular momentum through accretion, magnetic dipole radiation,
and GW radiation. Within this scenario, we calculate the evolution of GWs from a newborn magnetar
by considering the above three deformations. We find that the GW evolution depends on the physical
parameters of the magnetar (e.g., period and surface magnetic field), the adiabatic index, and the
fraction of poloidal magnetic energy to the total magnetic energy. In general the GW radiation from
a magnetically induced deformation is dominant if the surface magnetic field of the magnetar is
large, but the GW radiation from magnetar starquakes is more efficient when there is a larger
adiabatic index if all other magnetar parameters remain the same. We also find that the GW
radiation is not very sensitive to different magnetar equations of state.

	\end{abstract}
	
	\maketitle
	
	
	\section{introduction}
	Gravitational waves (GWs) are a prediction of the General Theory of Relativity. From a theoretical
	point of view, strong GW signals can be produced by cataclysmic events such as the merger of black
	holes (BHs), colliding neutron stars (NSs), as well as supernova explosions. In addition, weak GW
	signals are predicted to emanate from rotating NSs and are also predicted to be present in the Big
	Bang. In terms of observations, the first direct detection of GWs from a binary BH merger was the
	signal GW150914 observed by the Laser Interferometer Gravitational wave Observatory
	(LIGO; Ref~\cite{Abbott.et.al.2016}. Two years later, advanced LIGO and Virgo
	~\cite{Abbott.et.al.2017c,Abbott.et.al.2017B,Goldstein.et.al.2017,Pian.et.al.2017,Kasen.et.al.2017,
		Savchenko.et.al.2017,Zhang.et.al.2018} detected GW170817 from the merger of two neutron stars.
	Simultaneous to the GW170817 signal, electromagnetic signals emitted during this NS merger
	were detected, and the combined GW and EM signals opened the field of multimessenger astronomy to allow deeper
	exploration of the mysteries
	of the universe. However, the expected continuous GW signals emitted from isolated objects with
	asymmetric structures remain undetected thus far. These will be important scientific objects for
	next-generation GW detectors such as the Einstein telescope (ET).
	
	Newborn millisecond magnetars are promising candidate sources of continuous GW radiation
	~\cite{Kumar.et.al.2015,Lv.et.al.2017}. A direct search for postmerger GWs from the remnant of the
	binary NS merger GW170817 was performed by the aLIGO team ~\cite{Abbott.et.al.2017}, but no GW
	signals were found. Reference~\cite{Lv.et.al.2020} found indirect evidence of GW radiation in the
	afterglow of GRB 200219A, but the GW signal is too weak to be detected by aLIGO and Virgo.

	To date, many deformation mechanisms that enable an isolated system to emit GWs have been proposed
	in the literature~\cite{Johnson-McDaniel.et.al.2013,Giliberti.et.al.2021,Haskell.et.al.2006,
		Mastrano.et.al.2011,Zhong.et.al.2019,Ushomirsky.et.al.2000,Andersson.et.al.2003}. Magnetars are generally
	believed to have strong magnetic fields~\cite{Duncan.et.al.1992,Thompson.et.al.1993,Dai.et.al.1998,Dai.2004,
		Dai.et.al.2012,Lv.et.al.2014,lv.et.al.2015}, and the magnetic stress is too large for the magnetar to
	maintain a long-lasting spherical structure~\cite{Chandrasekhar.et.al.1953}. Magnetars with large toroidal
	magnetic fields tend to
	become powerful GW emitters~\cite{Cutler.2002}. The dynamical simulations of
	Ref.~\cite{Horowitz.et.al.2009} suggest that the NS crust is likely very strong and can support
	mountains large enough to produce GWs that can be detected in large-scale interferometers.
	The authors of Ref.~\cite{Gittins.et.al.2021} considered several examples for the form of the deforming force, and
	calculated that the maximum ellipticity that can support a neutron star crust is on the order of
	$10^{-8}-10^{-7}$. Afterwards, they applied it to the relativity case, and found that the maximum
	deformation that can support the crust of neutron star is two orders lower than the Newtonian
	case~\cite{Gittins.et.al.2021.2}. On the other hand, strong centrifugal forces in such magnetars
	can also break the NS's crust to result in starquakes, and form an asymmetric structure of the star
	~\cite{Giliberti.et.al.2021}. Moreover, magnetars can be born in the core collapse of a massive
	star~\cite{Wheeler.et.al.2000,Bucciantini.et.al.2009,Bucciantini.et.al.2008} or from the merger of
	binary stars~\cite{Rosswog.et.al.2003,Metzger.et.al.2008,Giacomazzo.et.al.2013,Yoon.et.al.2007}. A
	fraction of the remnant material ejected in these processes does not reach the escape velocity and
	falls back. During the accretion process, the magnetic poles of magnetars with high accretion rates
	will form significant accretion columns~\cite{Zhong.et.al.2019}. Further studies were carried
	out in Ref.~\cite{Sur.et.al.2021}, and they take into account the time variables of magnetar
	parameters, such as accretion rate, spin period, magnetic field, and moments of inertia. The
	evolution of inclination angle is also considered in their study, which found that the magnetic axis
	is orthogonal to the axis of rotation immediately after the birth of the star, which causes the
	accretion column to produce time-varying quadrupoles and GW radiation~\cite{Sur.et.al.2021}. In
	previous studies, several magnetar deformation mechanisms were even adopted to power the GW
	signals, but the authors did not consider simultaneously the contributions of all possible
	deformation mechanisms~\cite{Johnson-McDaniel.et.al.2013,Giliberti.et.al.2021,Haskell.et.al.2006,
		Mastrano.et.al.2011,Zhong.et.al.2019,Ushomirsky.et.al.2000,Andersson.et.al.2003}.
	
	One basic problem is to describe the evolution of GW radiation if the three lines of deformation
	mechanisms, i.e. magnetically induced deformation, starquake-induced ellipticity, and an accretion
	mountain are considered simultaneously. Which magnetar deformation mechanism dominates the
	contribution to the GW radiation? In this paper, we study the evolution of GWs emitted from newborn
	magnetars by considering the three most likely deformations above. In Sec.II, we introduce briefly
	the theory of GW radiation production with the above three lines of deformation mechanisms one by
	one. The calculation of the GW evolution with the period and surface magnetic field of the
	magnetar, adiabatic index, the fraction of poloidal magnetic energy, and for different NS equations
	of state are shown in Sec.III. Conclusions are drawn in Sec.IV with some additional discussion.

	\section{Deformation of the magnetar}
	In order to study the deformation of a magnetar, we first consider a background model of a
	spherical, nonmagnetic, nonrotating star. The hydrostatic equilibrium equation can be written as
	\begin{equation}
		\label{hydrostatic equilibrium equation}
		\nabla p_{0}+\rho_{0} \nabla \phi_{0}=0,
	\end{equation}
	where $p_{0}$ and $\rho_{0}$ are the initial pressure and density, respectively. The initial
	gravitational potential is $\phi_{0}$, and it obeys the Poisson equation:
	\begin{equation}
		\nabla^{2} \phi_{0}=4 \pi G \rho_{0},
	\end{equation}
	where $G$ is Newton's gravitational constant. The density configuration can be given as described in
	Refs.~\cite{Akgun.et.al.2008, Dall.et.al.2009} by adopting the polytropic equation of state (EOS)
	with $n=1$ (e.g., $p=k\rho^{1+1/n}$):
	\begin{equation}
		\rho_0=\frac{M}{4rR^{2}}\sin \left(\frac{\pi r}{R}\right),
	\end{equation}
	where $M$ and $R$ are the mass and radius of star, respectively.
	
	Here, we consider a newborn magnetar with three asymmetric structures, i.e., magnetically induced
	deformation under strong magnetic stress, deformation due to a series of high spin-induced
	starquakes, and asymmetric accretion columns caused by fallback accretion. The density distribution
	$\rho$ will be affected by the asymmetric perturbation of the starquakes
	$\rho_\mathrm{sta}^{\Delta}$, magnetic stress $\rho_\mathrm{mag}^{\Delta}$, and accretion
	$\rho_\mathrm{acc}^{\Delta}$. Moreover, the density distribution is also disturbed
	$\rho_\mathrm{cen}$ by the centrifugal force . The density distribution $\rho(\boldsymbol{r})$ can
	therefore be represented as:
	\begin{equation}
		\label{rho}		
		\rho(\boldsymbol{r})=\rho_0(r)+\rho_\mathrm{cen}(\boldsymbol{r})+\rho_\mathrm{sta}^{\Delta}(\boldsymbol{r})+\rho_\mathrm{mag}^{\Delta}(\boldsymbol{r})+\rho_\mathrm{acc}^{\Delta}(\boldsymbol{r}).
	\end{equation}
	
	The GW radiation of a magnetar is very sensitive to the ellipticity ($\epsilon$), which is defined
	as
	\begin{equation}
		\label{eq:inertia tensor}
		\epsilon=\frac{I_{yy}-I_{xx}}{I_{zz}}.
	\end{equation}
	Furthermore, as long as the difference of $I_{yy}-I_{xx}$ is small, the contribution of
	perturbation can be neglected in $I_{zz}$ to calculate the $\epsilon$. Hence, we use the moment of
	inertia of the spherical star $I_0$ instead of $I_{zz}$ in our calculations. Here, we adopt
	$I_{jk}$ to denote the component of the inertia tensor
	$\mathbf{I}$:
	\begin{equation}
		\label{eq:mathbfI}
		\mathbf{I}=\int \rho(\boldsymbol{r})\left(r^{2} \hat{\mathbf{I}}-\boldsymbol{r} \otimes
\boldsymbol{r}\right) dV,
	\end{equation}
	where $\hat{\mathbf{I}}$ is unit tensor, and $\boldsymbol{r} \otimes \boldsymbol{r}$ is the dyadic
	product of $\boldsymbol{r}$. One can substitute Eq.~(\ref{rho}) into Eq.~(\ref{eq:mathbfI}) to
	calculate the ellipticity. We find that the initial density $\rho_0$ and the perturbation of the
	density by the centrifugal force $\rho_\mathrm{cen}$ would provide the same contributions to all
	the components of the inertia tensor as in the case of uniform rotation. The terms cancel out in
	calculating the ellipticity, i.e. $I_{yy}=I_{xx}$. Hence, the total ellipticity $\epsilon$ is only
	dependent on the asymmetric deformation due to the starquake $\epsilon_{\mathrm{sta}}$, magnetic
	field $\epsilon_{\mathrm{mag}}$, and accretion $\epsilon_{\mathrm{acc}}$ terms. It can be expressed
	as
	\begin{eqnarray}
		\label{eq:ellipticity}
		{\epsilon}=&&I_{0}^{-1}
		\int_{V}\left[\rho_\mathrm{sta}^{\Delta}(\boldsymbol{r})+\rho_\mathrm{mag}^{\Delta}(\boldsymbol{r})+\rho_\mathrm{acc}^{\Delta}(\boldsymbol{r})\right]\left(r_{x}^2-r_{y}^2\right)d
		V \nonumber\\
		&&=\epsilon_{\mathrm{sta}}+\epsilon_{\mathrm{mag}}+\epsilon_{\mathrm{acc}}.
	\end{eqnarray}
	The asymmetric density will give the star a nonzero ellipticity that results in the production of
	GW radiation when it is rotated at the appropriate angle, the luminosity of which can be expressed
	as
	\begin{equation}
		\label{eq:Lgw}
		L_{\mathrm{gw}}=-\frac{2GI_{0}^{2}\Omega^{6}}{5c^{5}}\epsilon^{2}\sin^2\alpha(16\sin^2\alpha+\cos^2\alpha),
	\end{equation}
	where $\Omega$ is angular frequency, $c$ is the speed of light, and $\alpha$ is the misalignment
	angle. In Ref.~\cite{Sur.et.al.2021}, the authors found that the magnetic and rotational axes of a
	star would be orthogonal in the early stages ($ \leqslant 10$ ms), but
	gradually align after hundreds of years. These timescales are too short compared with the
	evolution time we are considering. Therefore, in our calculations we consider only the case
	that the magnetic axis ($z'$) is perpendicular to the rotation axis ($z$). One can estimate the
	upper limit of the GW radiation luminosity of the magnetar with $\alpha=90^\circ$. Within this
	scenario, all three deformations we consider will be maximized at the equator and the GW radiation
	will reach its maximum efficiency.

	\subsection{Magnetically induced deformation}
	For convenience, a spherical coordinate system ($r$, ${\theta}'$, ${\varphi}'$) with the magnetic
	axis ${z}'$ as the polar axis is used to calculate the deformation  caused by the magnetic
	field. Based on Eq.~(\ref{eq:ellipticity}), the magnetically induced ellipticity can be written
	as
	\begin{equation}
		\label{eq:epsilon_mag}
		\epsilon_{\mathrm{mag}}=I_{0}^{-1}
		\int_{V}\rho_\mathrm{mag}^{\Delta}(\boldsymbol{r})\left(r_{x}^2-r_{z'}^2\right)d V.
	\end{equation}
	Here, $\rho_\mathrm{mag}^{\Delta}$ is dependent on the configuration of the magnetic field. Let us
	adopt a universal configuration for the internal magnetic field with both poloidal
	$\boldsymbol{B}_{\mathrm{p}}=\left(B_{r},B_{\theta'},0\right)$ and toroidal components
	$\boldsymbol{B}_{\mathrm{t}}=\left(0, 0, B_{\phi'}\right)$ which satisfy the requirement of a
	stable magnetic field~\cite{Braithwaite.et.al.2006}. One can adopt a stream function $S(r,\theta')$
	to express each component of the magnetic field~\cite{Haskell.et.al.2008}
	\begin{equation}
		\label{eq:1}
		B_{r}=\frac{B_0\eta_p\partial_{\theta'} S}{2r^{2} \sin {\theta}'},
		B_{\theta'}=-\frac{B_0\eta_p\partial_r S}{2r \sin {\theta}'},
		B_{\phi'}=\frac{B_0\eta_t\beta(S)}{2r \sin {\theta}'},
	\end{equation}
	where $B_{0}$ is the strength of surface magnetic field at the dipole caps, and $\eta_{p}$ and
	$\eta_{t}$ are the relative strength of the poloidal and toroidal components, respectively. For a
	dipole magnetic field the stream function $S(r,\theta')$ can be written as
	\begin{eqnarray}
		\label{eq:2}
		S(r,\theta')=&&f(r)\sin^2\theta \nonumber\\
		&&=\frac{35}{8}\left(\frac{r^2}{R^2}-\frac{6r^4}{5R^4}+\frac{3r^6}{7R^6}\right) \sin^2\theta.
	\end{eqnarray}
	In order to ensure continuity of the magnetic field from internal to external across the
	surface of the star, we also adopt the same expression for $f(r)$ as in
	Ref~.\cite{Marchant.et.al.2011}. Furthermore, following the result from
	Ref.~\cite{Mastrano.et.al.2011}, the toroidal component should be limited within the region
	$S\geqslant 1$, so $\beta(S)$ can be defined as
	\begin{equation}
		\label{eq:3}
		\beta(S)=\left\{\begin{array}{ll}
			(S-1)^{2} & , S \geqslant 1 \\
			0 & , S<1
		\end{array}\right.
	\end{equation}
	Combining with Eqs.~(\ref{eq:2}),(\ref{eq:3}), and (\ref{eq:1}), one can derive the expression for the
	magnetic field configuration,
	\begin{equation}
		\label{eq:2.4}
		\mathbf{B}=B_0\left(\frac{f\eta_{p}\cos{\theta}'}{r^2},\frac{{f}'\eta_{p}\sin{\theta}'}{2r},\frac{\beta\eta_{t}}{2r\sin{\theta}'}\right).
	\end{equation}

	Assuming that the asymmetric density of the magnetic field $\rho_\mathrm{mag}^{\Delta}$ is small enough
by
	considering it as a perturbation $\delta \rho_\mathrm{mag}$ on a spherical background in
	Eq.~(\ref{hydrostatic equilibrium equation}), the first-order approximation of the perturbed
	momentum equation can be written as
	\begin{equation}
		\label{eq:2.5}
		\nabla \delta p_{\mathrm{mag}}+ \delta\rho_{\mathrm{mag}} \nabla \phi_0=\frac{\pounds}{\mu_0}.
	\end{equation}
	Here, we adopt the Cowling approximation which ignores the contribution of the perturbation of the
	gravitational potential and $\pounds=(\nabla \times \mathbf{B}) \times \mathbf{B}$ is Lorentz
	force. Combining with Eq.~(\ref{eq:2.4}) and Eq.~(\ref{eq:2.5}), one can obtain the solution to the
	perturbation of the density distribution, which is given as
	\begin{eqnarray}
		{\frac{d \phi_0}{d r}}\frac{R^2}{B_{0}^{2}}\mu_{0}
		\delta\rho_{\mathrm{mag}}=&&\frac{105r}{4}S+\frac{1-\Lambda}{8q\Lambda}\left(\frac{5r^3}{R^2}-7r\right)\nonumber\\
		&&\times\left(S^3-\frac{9S^2}{2}+9S-3\ln S-\frac{18}{11}\right)
		\label{eq:drho}
	\end{eqnarray}
	Following Ref.~\cite{Mastrano.et.al.2011}, we adopt the typical values of $\eta_p=1$ and
	$q=1.95\times 10^{-6}$. The ratio of the magnetic energy of the poloidal field to the total
	magnetic energy is $\Lambda=\eta_p^2/\left(\eta_p^2+q\eta_t^2\right)$. However, in reality
	very little is known about the value of the parameter $\Lambda$. Within the Newtonian magnetohydrodynamics (MHD) simulation,
	Ref.~\cite{sur.et.al.2020} found that all the initial configurations that they selected were
	unstable, and the ratio of the poloidal-toroidal energies became approximately stable when the
	instability developed on the order of an $\mathrm{Alfv\acute{e}n}$ crossing timescale. If this is
	the case, the poloidal component will contribute $\ge80\%$ of total magnetic energy (i.e.,
	$\Lambda\ge 0.8$). On the contrary, within the relativistic MHD simulation,
	Ref.~\cite{sur.et.al.2022} shows that the $\Lambda$ would stabilize at an equilibrium value of 0.2
	when the toroidal initial setup dominated. The timescale from instability to approximately
	stability ($<1$ s) is much shorter than the timescale we consider, so we ignore the evolution of
	$\Lambda$ in the following calculations. The first and second terms on the right-hand side of
	Eq.~(\ref{eq:drho}) describe the contributions of the poloidal and toroidal components of the
	magnetic field to the density perturbations, respectively. Due to the spherical symmetry of the
	density perturbation, we keep only the spheroidal terms and hence Eq.~(\ref{eq:epsilon_mag}) can be
	rewritten as~\cite{Mastrano.et.al.2011}
	\begin{eqnarray}
		\label{eq:epsilon_mag2}
		\epsilon_{\mathrm{mag}}=&&\pi  I_{0}^{-1} \int_{V}\delta \rho_{\mathrm{mag}}(r, {\theta}') r^{4}
\sin
		{\theta}'\left(1-3 \cos ^{2} {\theta}'\right)d r d
		{\theta}'\nonumber\\
		&&=6.262 \times 10^{-6}\left(\frac{B_{0}}{10^{15} \mathrm{~G}}\right)^{2}\left(\frac{M}{1.4
			M_{\odot}}\right)^{-2}\nonumber\\
		&&\times\left(\frac{R}{10^{6} \mathrm{~cm}}\right)^{4}\left(1-\frac{0.385}{\Lambda}\right).
	\end{eqnarray}

	\subsection{Starquake-induced ellipticity}
	A neutron star is usually considered to consist of a fluid core of radius ($r_c$) and an elastic
	crust. Initially, the crust of a neutron star would form without strain when the star is born
	by rapidly rotating. However, the stress would build up in the crust until a breaking condition was
	reached when the rotation rate was changed. The breaking condition is evaluated by the Tresca criterion:
	the strain angle $\alpha_s$ is half of the breaking strain:
	$\sigma_{\mathrm{max}}$~\cite{Christensen.2013}
	\begin{equation}
		\alpha_s=\frac{\sigma_{\mathrm{max}}}{2}.
	\end{equation}	
	It is worth noting that the value of $\sigma_{\mathrm{max}}$ is very uncertain.
	Reference \cite{Giliberti.et.al.2021} adopted a larger breaking strain $\sigma_{\mathrm{max}}=10^{-1}$,
	and found that the breaking frequency is in the range $200-600$ Hz for a typical $M = 1.4M_{\odot
	}$ of NS, while $\sigma_{\mathrm{max}}=10^{-5}$ is adopted by~\cite{Ruderman.1991}, and found that
	the fracture frequency would be about two orders of magnitude smaller than that of $200-600$ Hz. It
	means that the crust of the neutron star will be fractured when we change the frequency a little bit (a few Hz). 
	Therefore, the local crust at the equator undergoes a sufficient number of breaking events
	during spin evolution of the neutron star. Because of the different shear modulus between the fluid
	configurations and elastic crust, the cumulative effect of a series of starquakes makes the elastic
	crust tend to the equilibrium structure of fluid configuration and produces a large ellipticity of
	the star.
	
	It is difficult to calculate exactly the density changes caused by starquakes
	$\rho_\mathrm{sta}^{\Delta}$, but we can estimate roughly the maximum ellipticity by comparing the
	difference in the moment of inertia between the two different configurations,
	\begin{equation}
		\label{eq:epsilon_sta}
		\epsilon_{\mathrm{sta} }(t)=\frac{I_{yy}^F-I_{xx}^E}{I_0}.
	\end{equation}
	Here, the superscripts $E$ and $F$ are the elasticity and fluid configurations, respectively.
	We cannot ensure that each starquake will release all the stresses of the crust and result in the
	crust reaching the fluid configuration instantaneously. Hence, the ellipticity in our calculations
	should be an upper limit.
	
	We assume that the crust is formed when the star is rotating rapidly at the initial angular
	frequency ($\Omega_0$). The contribution of density perturbations from starquakes
	($\rho_{\mathrm{sta}}^\Delta$), magnetic fields ($\rho_{\mathrm{mag}}^\Delta$), and accretion
	($\rho_{\mathrm{acc}}^\Delta$), is much smaller than the contribution of the millisecond rotating
	centrifugal force to density perturbation ($\rho_\mathrm{cen}$). Therefore, the density distribution at
	this time can be approximated by
	\begin{eqnarray}	
		\rho_\mathrm{ini}(\boldsymbol{r})&&=\rho_0(r)+\rho_\mathrm{cen}(\boldsymbol{r})+\rho_\mathrm{sta}^{\Delta}(\boldsymbol{r})+\rho_\mathrm{mag}^{\Delta}(\boldsymbol{r})+\rho_\mathrm{acc}^{\Delta}(\boldsymbol{r})
		\nonumber\\
		&&\approx\rho_0(r)+\rho_\mathrm{cen}(\boldsymbol{r}).
	\end{eqnarray}
	If this is the case, the strain would build up with the change of angular frequency $\Omega$,
	and the inconsistency of shear modulus between the elastic crust and fluid core would cause the 
	density to have a different response to the change of centrifugal force:
	\begin{equation}
		\label{eq:rho}
		\rho(\boldsymbol{r})=\rho_\mathrm{ini}(\boldsymbol{r})+\rho_\mathrm{cen}^{\Delta}(\boldsymbol{r}),
	\end{equation}
	Here, $\rho_\mathrm{cen}^{\Delta}$ represents the perturbation of density distribution caused by
	the change of angular frequency $\Omega-\Omega_0$.
	
	Combining with Eq.~(\ref{eq:rho}) and Eq.~(\ref{eq:mathbfI}), one can find the expression for the
	inertia tensor to be
	\begin{equation}
		\mathbf{I}=\mathbf{I_0}+\mathbf{I}_\mathrm{cen}+\mathbf{I}^{\Delta}_\mathrm{cen},
	\end{equation}
	where $\mathbf{I_0}$ is the undisturbed inertial tensor, $\mathbf{I}_\mathrm{cen}$ is the
	change of inertial tensor cause by the initial angular frequency $\Omega_0$, and
	$\mathbf{I}^{\Delta}_\mathrm{cen}$ is the perturbation in inertial tensor caused by the change of
	angular frequency $\Omega-\Omega_0$. We noticed that $\mathbf{I_0}$ and $\mathbf{I}_\mathrm{cen}$
	give the same contribution to the $I_{yy}$ and $I_{xx}$ components of the inertia tensor
	$\mathbf{I}$, so that, they will be canceled out when we calculate the $\epsilon_{\mathrm{sta}}$.
	Therefore, Eq.~(\ref{eq:epsilon_sta}) can be rewritten as
	\begin{equation}
		\label{eq:epsilon_sta3}
		\epsilon_{\mathrm{sta} }(t)=\frac{I^{\Delta F}_{yy}-I^{\Delta E}_{xx}}{I_0},
	\end{equation}
	where $I^{\Delta F}_{yy}$ and $I^{\Delta E}_{xx}$ are the components of
	$\mathbf{I}_\mathrm{cen}^{\Delta}$.
	
	Moreover, it should be noted that the initial configuration 
	$\rho_\mathrm{ini}(\boldsymbol{r})\approx\rho_0(r)+\rho_\mathrm{cen}(\boldsymbol{r})$ is not
	spherical and different from the spherical configuration $\rho_\mathrm{ini}(\boldsymbol{r})=\rho_0$
	in Ref.~\cite{Giliberti.et.al.2021}, despite that the initial configuration deviation caused by
	centrifugal force $\rho_\mathrm{cen}(\boldsymbol{r})$ is axisymmetric and does not directly
	contribute to the ellipticity. The centrifugal perturbation
	$\rho_\mathrm{cen}^{\Delta}(\boldsymbol{r})$ still has an effect on some unknowns of the problem
	due to the difference of initial configuration, though $\rho_\mathrm{cen}(\boldsymbol{r})$ may be a
	small fraction of $\rho_0(r)$. However, to calculate $\rho_\mathrm{ini}(\boldsymbol{r})$ precisely
	is indeed a difficult and complex problem with many uncertainties; we will still use the
	unstrained spherical configuration mentioned in Ref.~\cite{Giliberti.et.al.2021} to get the
	estimate of $\mathbf{I}_\mathrm{cen}^{\Delta}$.		
	By selecting the reference frame of the rotation axis $z$ as the polar axis (i.e., $\theta=0$ at
	the rotation axis $z$), $\mathbf{I}_\mathrm{cen}^{\Delta}$ can be divided into two terms with
	$\mathbf{I}_{00}^{\Delta}$ and $\mathbf{I}_{20}^{\Delta}$,
	\begin{equation}
		\label{eq:Delta I_00}
		\mathbf{I}_{00}^{\Delta}=\frac{8 \pi}{3} Diag[1,1,1] \int_{0}^{a} \rho_{00}^{\Delta}(r) r^{4} d r,
	\end{equation}
	\begin{equation}
		\label{eq:Delta I_ij}
		\mathbf{I}_{20}^{\Delta}=\frac{4 \pi}{5}Diag[\frac{1}{3},\frac{1}{3},\frac{2}{3}]\int_{0}^{a}
		\rho_{20}^{\Delta}(r) r^{4} d r,
	\end{equation}
	where each $\rho_{\ell m}^{\Delta}$ is the coefficient of the spherical harmonic expansion of
	$\rho_\mathrm{cen}^{\Delta}$ with degree $\ell$ and order $m$. The spherical harmonic expansion of
	$\rho_\mathrm{cen}^{\Delta}$ is independent of ${\varphi}'$ and the only remaining terms are the
	$l=0$ and $l=2$ (\cite{Sabadini.et.al.2016,Giliberti.et.al.2021}). We find that the
	$\mathbf{I}_{00}^{\Delta}$ also give the same contribution to all the components of the inertia
	tensor $\mathbf{I}$, and it will be canceled out as well when the $\epsilon_{\mathrm{sta}}$ is
	calculated. Therefore, we only focus on $\mathbf{I}_{20}^{\Delta}$ and represent it as
	\begin{equation}
		\mathbf{I}_{20}^{\Delta}=\Delta\mathbf{I}.
	\end{equation}
	Here, caution is needed. For a pure fluid and elastic configuration, the $l=2$, $m=0$
	deformation $\rho_{20}^{\Delta}$ is axisymmetric if the centrifugal force distortion axis is
	aligned with the rotation axis, and does not lead to GW emission. However, the postquakes
	configuration $\Delta\mathbf{I}^Q$ will be range in the elastic configuration and the fluid
	configuration, and the components have $\Delta I^{Q}_{yy}\le\Delta I^{E}_{yy}$ and $\Delta
	I^{F}_{xx}\le\Delta I^{Q}_{xx}$. Therefore, one can approximately present the upper limit of 
	nonaxisymmetric configurations by comparing the principal moment of inertia of the two axisymmetric
	configurations with $l=2$,$m=0$ harmonics.		
	On the other hand, the total perturbed potential $\Phi^{\Delta}$ can also be expanded in spherical
	harmonics as Ref.~\cite{Chao.et.al.1987},
	\begin{eqnarray}
		\label{eq:Phi}
		\Phi^{\Delta}(r)=&&\sum_{\ell=0}^{\infty}\sum_{m=-\ell}^{\ell}\Phi^{\Delta}_{\ell m}(r)Y_{\ell 
			m}(\theta,\varphi)\nonumber\\
		&&=-\sum_{\ell=0}^{\infty}\sum_{m=-\ell}^{\ell} \frac{4 \pi G r}{(2 \ell+1)}Y_{\ell m}(\theta,
		\varphi)\nonumber\\
		&&\times\int_{0}^{a} \rho_{\ell m}^{\Delta}
		\left(r^{\prime}\right)\left(\frac{r^{\prime}}{r}\right)^{\ell+2}d r^{\prime}.
	\end{eqnarray}
	The spherical harmonic expansion of $\Phi^{\Delta}$ is independent of ${\varphi}'$, and we
	adopt one term with $l=2,m=0$. One has
	\begin{equation}
		\label{eq:Phi_2m}
		\Phi_{2 0}^{\Delta}(r)=-\frac{4 \pi G}{5 r^{3}} \int_{0}^{a} 
		\rho_{20}^{\Delta}\left(r^{\prime}\right)r^{\prime 4} d r^{\prime}.
	\end{equation}
	Comparing to Eq.(\ref{eq:Delta I_ij}), one obtains the MacCullagh formula:
	\begin{equation}
		\label{eq:DeltaI}
		\Delta\mathbf{I}=-\mathrm{Diag}[\frac{1}{3},\frac{1}{3},\frac{2}{3}]\frac{R^{3}}{G} \Phi_{2
			0}^{\Delta}(R).
	\end{equation}
	Based on Eq.(\ref{eq:DeltaI}), it is not necessary to solve the density perturbation to calculate 
	the change in the moment of inertia, but only to invoke the perturbation of gravitational potential 
	from the centrifugal force. Therefore, Eq.(\ref{eq:epsilon_sta}) can be written as
	\begin{equation}
		\label{eq:epsilon_sta2}
		\epsilon_{\mathrm{sta} }(t)=\frac{R^{3}}{3 I_{0} G}\left[\Phi_{20}^{\Delta E}(R)-\Phi_{20}^{\Delta
			F}(R)\right],
	\end{equation}
	where $\Phi_{20}^{\Delta F}(R)$ and $\Phi_{20}^{\Delta E}(R)$ are the total perturbation of the
	surface potential of the star for the elastic and fluid configurations, respectively. Following the
	method of Ref.~\cite{Giliberti.et.al.2021}, the internal fluid and external crust of the star
	correspond to different shear moduli, respectively. The shear modulus of the elastic external
	configuration can be described by the typical formula found in~\cite{Cutler.et.al.2003}
	\begin{equation}
		\mu=10^{-2} P.
	\end{equation}
	Moreover, the high temperature of the elastic crust is more fluidlike and has a lower shear modulus
	than that of cold crust~\cite{Hoffman.et.al.2012}.
	
	On the other hand, in order to describe the structure of the NS, one needs to present the bulk modulus
	$\kappa$ which is proposed in Refs.\cite{Giliberti.et.al.2019,Giliberti.et.al.2021}:
	\begin{equation}
		\kappa(r)=\gamma P(r),
	\end{equation}
	where $\gamma$ is the adiabatic index. It describes the response time to reach
	thermodynamic equilibrium in the star when the perturbation interacts with the matter. If one system
	completely reaches thermodynamic equilibrium when the dynamical timescale of the perturbation is
	longer than the timescale of a chemical reaction, it is called slow dynamics. If this is the case,
	$\gamma$ is the equilibrium adiabatic index, $\gamma=\gamma_{eq}=(n+1)/n=2$ for the $n=1$ polytropic
	EOS.

	\subsection{Accretion mountain}
	A magnetar may survive a double-neutron star merger or massive star collapse.
	Reference~\cite{Sur.et.al.2021} first considered the available mass from a merger as opposed to a
	massive star collapse. The matter surrounding the magnetar will fall back along the magnetic field
	lines to the dipole caps of the star and an accretion column may form as time goes on. Here, we
	adopt a parametrization of the accretion rate which is presented in Ref.~\cite{Piro.et.al.2011}:
	\begin{equation}
		\dot{M}\approx \dot{M}_{\text {early }}=\eta 10^{-3} t^{1 / 2} M_{\odot}~\mathrm{s}^{-1},
	\end{equation}
	where $\eta\approx 0.1-10$ accounts for different explosion energies
	~\cite{MacFadyen.et.al.2001,Zhang.et.al.2008,Piro.et.al.2011}. The total mass of the star is
	increased by accretion, and it is a function of time~\cite{Sur.et.al.2021}:
	\begin{equation}
		M_{\mathrm{tol}}(t) =M_{0}+\int_{0}^{t} \dot{M} d t=M_{0}+\frac{2}{3} \eta 10^{-3} t^{3 / 2}
	\end{equation}

	Following the derivation in Ref.~\cite{Zhong.et.al.2019}, the mass of matter that falls back onto
	the poles of the magnetar can be approximated as
	\begin{eqnarray}
		{M_\mathrm{acc}}(t)=&&8.83 \times 10^{-8}\eta^{3/28} t^{3/56}M_{\odot}\nonumber\\
		&&\times
		\left(\frac{M}{1.4M_{\odot}}\right)^{\frac{25}{56}}\left(\frac{B_0}{10^{15}\mathrm{~G}}\right)^{-\frac{5}{7}}
		\left ( \frac{R}{10^6 \mathrm{~cm}}\right
		)^{\frac{125}{56}}.
	\end{eqnarray}
	In fact, there is a small fraction of fallback matter that can reach the surface of the star 
	and will affect the star's structure.  However, the effect is too weak and can be ignored
	in comparison to the contributions of starquakes, magnetic forces, and the accretion column
	itself. The size of accretion mountain is small enough in comparison with the radius of the 
	magnetar that we can treat it as a point mass in calculating the ellipticity,
	\begin{equation}
		\label{eq:epsilon_acc}
		\epsilon_{\mathrm{acc}}(t)\approx \frac{I_0-\left(I_0+2M_\mathrm{acc}R^2\right)}{I_0}=-2 M_{acc}
		R^{2}I_0^{-1}.
	\end{equation}
	Here, the negative ellipticity means that the deformation of the accretion column is perpendicular
	to magnetically induced deformation. In our calculations, we do not consider the evolution of
	parameters (e.g., moments of inertia, magnetic field, and inclination angle) which have been
	studied in Ref.~\cite{Sur.et.al.2021}.

	\section{GW radiation evolution in a magnetar}
	Based on Eq.(\ref{eq:Lgw}), the GW luminosity has a very sensitive dependence on the angular
	frequency $\Omega$ ($L_{gw}\propto \Omega^6$). Let us first consider the evolution of the angular
	frequency caused by magnetic dipole radiation, GW radiation, and accretion before
	calculating the GW luminosity of a deformed magnetar. As mentioned in Ref.~\cite{Piro.et.al.2011}, 
	there are two important radii for a magnetar in the propeller regime: the
	Alfv{\'e}n radius ($r_m$) and the corotation radius ($r_c$). They are defined as
	\begin{equation}
		r_m=\left(\frac{B_0^4R^{12}}{GM\dot{M}^2}\right)^{1/7},r_c = \left(\frac{GM}{\Omega^2}\right)^{1/3}.
	\end{equation}
	The accretion torque on the star is given by Ref.~\cite{Piro.et.al.2011}
	\begin{equation}
		N_{\mathrm{acc}}=
		\begin{cases}
			&\left(1-\Omega / \Omega_{\mathrm{k}}\right)(G M R)^{1 / 2} \dot{M} \quad \text{ if } r_{m}<R \\
			&\xi\left(G M r_{m}\right)^{1/2} \dot{M} \quad ~~~~~~~\text{ if } r_{m}>R
		\end{cases},
	\end{equation}
	where  $\Omega_k = (GM/R^3)^{1/2}$ is the break-up frequency, and $\xi=1-(r_m/r_c)^{3/2}$ is a
	dimensionless parameter. A positive $\xi$ indicates the star is spun up by accretion and a negative
	$\xi$ indicates the star is spun down through the expulsion of matter~\cite{Piro.et.al.2011}. If
	the magnetar is formed from the collapse of a massive star, the mass of the accretion disk is
	required to be less than $ 1M_{\odot}$~\cite{Meszaros.2006}. However, if the magnetar is formed
	from a binary NS merger, the ejecta mass is expected to be much lower than
	$0.2M_{\odot}$~\cite{Radice.et.al.2018,Bernuzzi.et.al.2020,Bernuzzi.2020,Radice.et.al.2020}. Here,
	we adopt $\eta=10$ and a mass of 0.2$M_{\odot}$ available for accretion as
	Ref.~\cite{Sur.et.al.2021} does. If this is the case, the effects of accretion will eventually
	disappear when the ejected material from the merger of the binary neutron stars is exhausted via
	accretion,
	\begin{equation}
		N_{\mathrm{acc}}=0, ~\text{ if } \int_{0}^{t} \dot{M} dt>0.2M_{\odot}.
	\end{equation}

	In addition, the magnetar loses angular momentum due to negative torques generated by
	magnetic dipole radiation
	\begin{equation}
		N_{\mathrm{dip}}=-\frac{B_0^2R^6\Omega^3}{6c^3}
	\end{equation}
	and GW radiation
	\begin{equation}
		N_{\mathrm{gw}}=-\frac{L_{\mathrm{gw}}}{\Omega}.
	\end{equation}
	Following the method of Ref.~\cite{Piro.et.al.2011}, we ignore the contribution from
	neutrino-induced spin-down. Hence, the spin evolution can be given as
	\begin{equation}
		\label{eq:domega}
		\frac{\mathrm{d} \Omega }{\mathrm{d}
			t}=\frac{N_{\mathrm{tol}}}{I_0}=\frac{N_{\mathrm{acc}}+N_{\mathrm{dip}}+N_{\mathrm{gw}}}{I_0},
	\end{equation}
	
	\begin{figure}
		\includegraphics[width=3.5in]{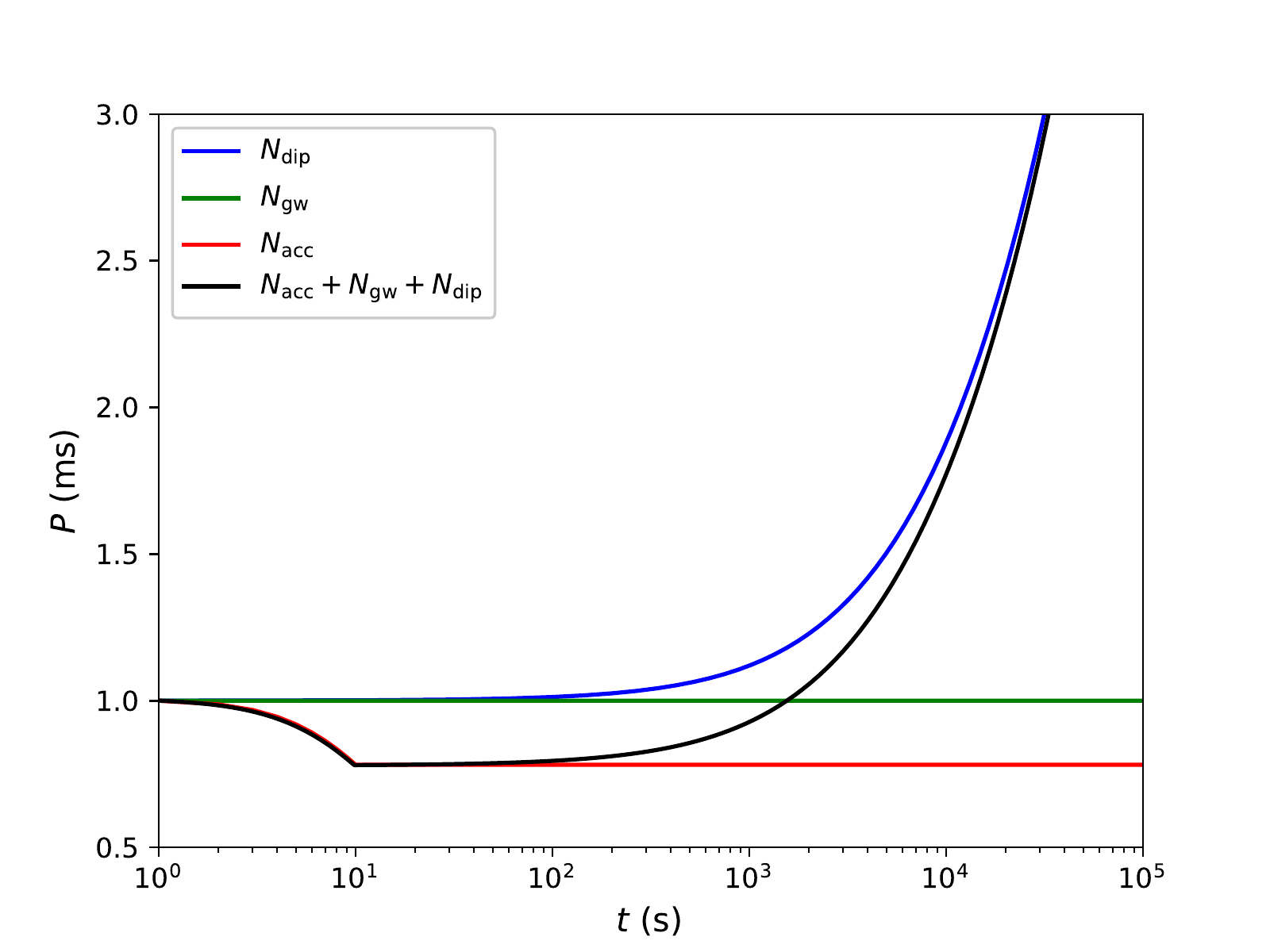}
		\caption{\label{fig:Omega_t}Period of a magnetar as function of time for given Skyrme Lyon (SLy) 
			EOS, $B_0=10^{15}\mathrm{G}$, and $\Lambda=0.2$. The solid red, blue, 
			and green lines are the accretion torque, GW radiation torque, and dipole radiation torque, 
			respectively. The black curve is the total contribution from all three.}
	\end{figure}
	
	Figure~\ref{fig:Omega_t} shows the evolution of the period for given EOS of the NS, strength of the
	magnetic field, and fraction of poloidal magnetic energy. We find that the accretion torque
	$N_{\mathrm{acc}}$ is dominant during the spin-up of the magnetar at an early time ($t\lesssim 10
	$ s) before the available accreted mass is exhausted. At a later time, the angular momentum
	of the magnetar is carried away mainly via magnetic dipole radiation. The contribution of the GW
	radiation is small and can be ignored.

	As mentioned in Ref.~\cite{Giliberti.et.al.2019}, the total perturbation potential is dependent on
	the angular frequency of the magnetar. Based on Eq.(\ref{eq:domega}), one can easily obtain the
	evolution of the total perturbation potential with time, or the evolution of
	$\epsilon_{\mathrm{sta}}$. Combining with Eqs.~(\ref{eq:Lgw}), (\ref{eq:epsilon_mag2}), 
	(\ref{eq:epsilon_sta2}), (\ref{eq:epsilon_acc}), and (\ref{eq:domega}),
	one can obtain the evolution of total GW luminosity, which is written as
	\begin{equation}
		\label{eq:L}
		L_{\mathrm{gw}}(t)=-\frac{32}{5}
		\frac{GI_{0}^{2}}{c^{5}}\left[\epsilon_\mathrm{acc}(t)+\epsilon_\mathrm{mag}+\epsilon_\mathrm{sta}(t)\right]^{2}
		\Omega(t)^{6}.
	\end{equation}

\begin{figure*}
		\includegraphics[width=7in]{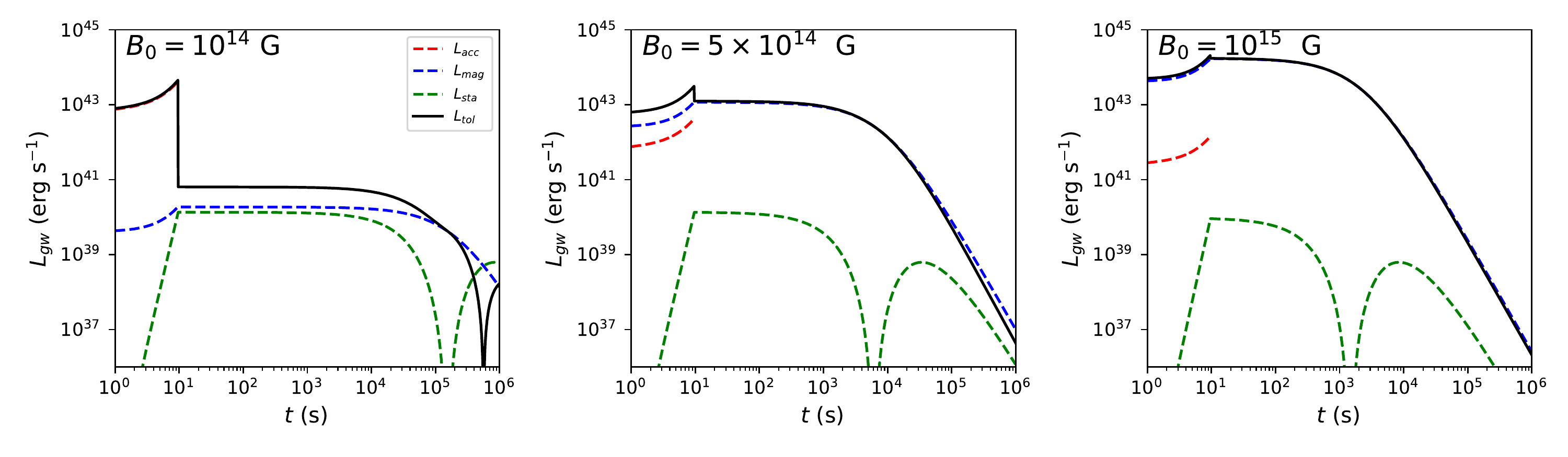}
		\caption{\label{fig:Lgw_B0}Evolution of GW radiation with fixed $\Lambda=0.2$ and given SLy EOS for
			different magnetic field strengths ($B_0=10^{14}\mathrm{G}, 5\times10^{14}\mathrm{G},
			10^{15}\mathrm{G}$).}
	\end{figure*}

	Figure~\ref{fig:Lgw_B0} shows the evolution of GW luminosity with different magnetic field strengths
	(e.g., $B_0=10^{14}\mathrm{G}, 5\times10^{14}\mathrm{G}, 10^{15}\mathrm{G}$) with fixed
	$\Lambda=0.2$ and a given EOS. We find that the GW radiation caused by an accretion mountain is
	stronger than the GW radiation from starquake-induced deformation or magnetically induced
	deformation for $B_0=10^{14}\mathrm{G}$. Because of the hypothesis above that the crust breaking
	occurred at the equator and the definition of starquake-induced ellipticity in
	Eq.~(\ref{eq:epsilon_sta3}), the $\epsilon_{\rm sta}$ will be negative or positive for
	$\Omega-\Omega_0>0$ and $\Omega-\Omega_0<0$, respectively. According to Eq.~(\ref{eq:epsilon_mag2}),
	$\epsilon_{\rm mag}$ is dependent on the value of $\Lambda$, and $\epsilon_{\rm mag}$ is negative
	and in the same direction as $\epsilon_{\rm acc}$ if $\Lambda<0.385$, and vice versa. Thus, the
	positive ellipticity of magnetars caused by starquakes and the negative ellipticity of magnetars
	caused by a larger toroidal magnetic field will cancel each other out when $\Omega<\Omega_0$, and
	the total GW radiation will decrease afterward or even disappear when 
	$\epsilon_{\rm sta}=-\epsilon_{\rm mag}$.
	
	As the strength of the magnetic field increases $(B_0=5\times10^{14}\mathrm{G})$, the GW radiation
	caused by magnetically induced deformation becomes increasingly strong. $L_{\mathrm{sta}}$
	will increase gradually during accretion phase because of increased $\epsilon_{\rm sta}$ when the star
	spins up. As well, $\epsilon_{\rm sta}$ will gradually decrease or even disappear completely when
	the accretion stops. When the star spins down, $\epsilon_{\rm sta}$ becomes negative and goes on
	decreasing, hence $L_{\mathrm{sta}}$ ($\propto \epsilon^2_{\rm sta}$) can be increased again. When
	the strength of the magnetic field is strong enough, i.e. reaching $10^{15}\mathrm{G}$, the GW
	radiation caused by the magnetically induced deformation is dominant.

\begin{figure*}
		\includegraphics[width=7in]{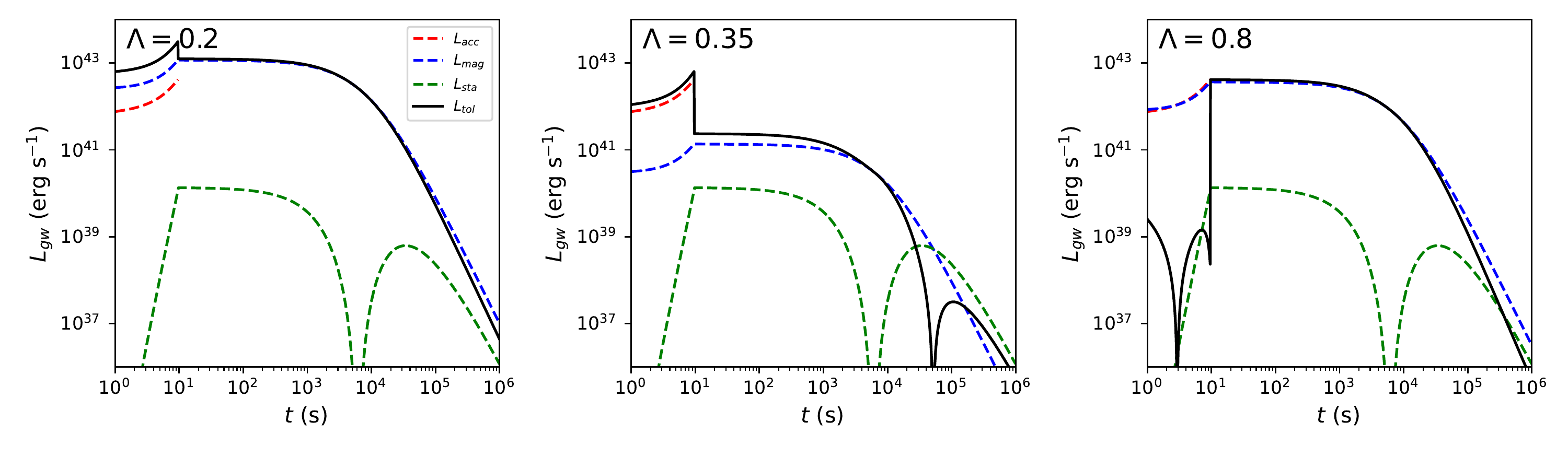}
		\caption{\label{fig:Lgw_A}Evolution of GW radiation with fixed $B_0=5\times10^{14}\mathrm{G}$ and
			given SLy EOS for different poloidal magnetic field components ($\Lambda=0.2$, 0.35, 0.8).}
	\end{figure*}

	In order to test the dependence of the GW radiation on $\Lambda$ with a fixed-strength magnetic
	field ($B_0=5\times10^{14}\mathrm{G}$) and given EOS, we present the evolution of GW radiation with
	different $\Lambda$ in Fig.~\ref{fig:Lgw_A}. From the MHD simulation point of view, for given
	different initial magnetic field configurations, $\Lambda$ is as large as 0.8 in a relativistic
	situation, but in a Newtonian simulation, the $\Lambda$ would stabilize at an equilibrium value of
	0.2~\cite{sur.et.al.2020,sur.et.al.2022}. So, we choose $\Lambda=0.2, 0.35, 0.8$ in our
	calculations. With a large toroidal component of the magnetic field $\Lambda=0.2$, the
	contributions to the GW radiation mainly come from $L_{\rm{mag}}$. Decreasing the fraction of
	the toroidal component (increasing the value of $\Lambda=0.385$) will make the contribution of
	$L_{\rm mag}$ gradually decrease, and $L_{\rm sta}$ will dominate at later time. For
	$\Lambda>0.385$, the total GW radiation is dominated by magnetically induced deformation with
	$\epsilon_{\rm mag}>0$. Therefore, the $L_{\rm tol}$ has an initially sharp drop with $\Lambda=0.8$
	due to the offset of $\epsilon_{\rm mag}$ and $-\epsilon_{\rm acc}$.

\begin{figure*}
		\includegraphics[width=7in]{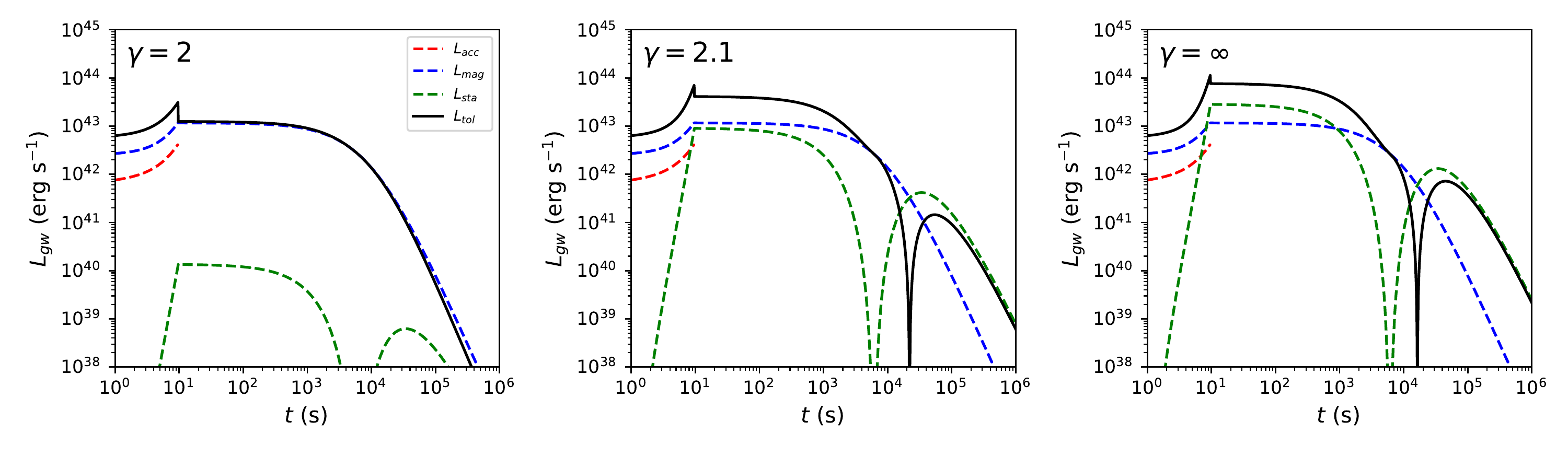}
		\caption{\label{fig:Lgw_gamma} Evolution of GW radiation calculated with different adiabatic
			indices ($\gamma=2, 2.1, \infty$) for SLy EOS and fixed $B_0=10^{15}$, $\Lambda=0.2$.}
	\end{figure*}

	Similarly with Fig.~\ref{fig:Lgw_B0} and Fig.~\ref{fig:Lgw_A}, we also present the evolution of the
	GW radiation with different adiabatic indices ($\gamma=2, 2.1, \infty$) for SLy EOS and fixed
	$B_0=10^{15}$, $\Lambda=0.2$ in Fig.~\ref{fig:Lgw_gamma}. The response timescale of the density
	perturbation is approximately equal to the timescale for the system reaching complete thermodynamic
	equilibrium when the system is perturbed, and it is dependent on the adiabatic index $\gamma$. A
	larger $\gamma$ corresponds to a shorter response timescale of the density perturbation.
	Figure~\ref{fig:Lgw_gamma} shows the results for different adiabatic indices ($\gamma=2, 2.1,
	\infty$). We find that GW radiation from magnetically induced deformation is dominant with
	$\gamma=2$, but when we increase $\gamma$ from 2 to 2.1, the GW radiation of the starquake-induced
	deformation strengthens significantly and becomes comparable with $L_\mathrm{mag}$. A similar
	result is also found if we increase $\gamma$ to infinity (see also
	Ref.~\cite{Giliberti.et.al.2021}). For $\gamma=2.1$ or even infinity, the total GW luminosity
	rapidly decreases when the magnetar is spun down. At some point, however, $\epsilon_\mathrm{sta}$
	supported by the centrifugal force perturbation will gradually become stronger, and
	will decrease until it is comparable with $-\epsilon_\mathrm{mag}$. After that, the total GW
	radiation is dominated by the starquake-induced deformation.

\begin{figure*}
		\includegraphics[width=7in]{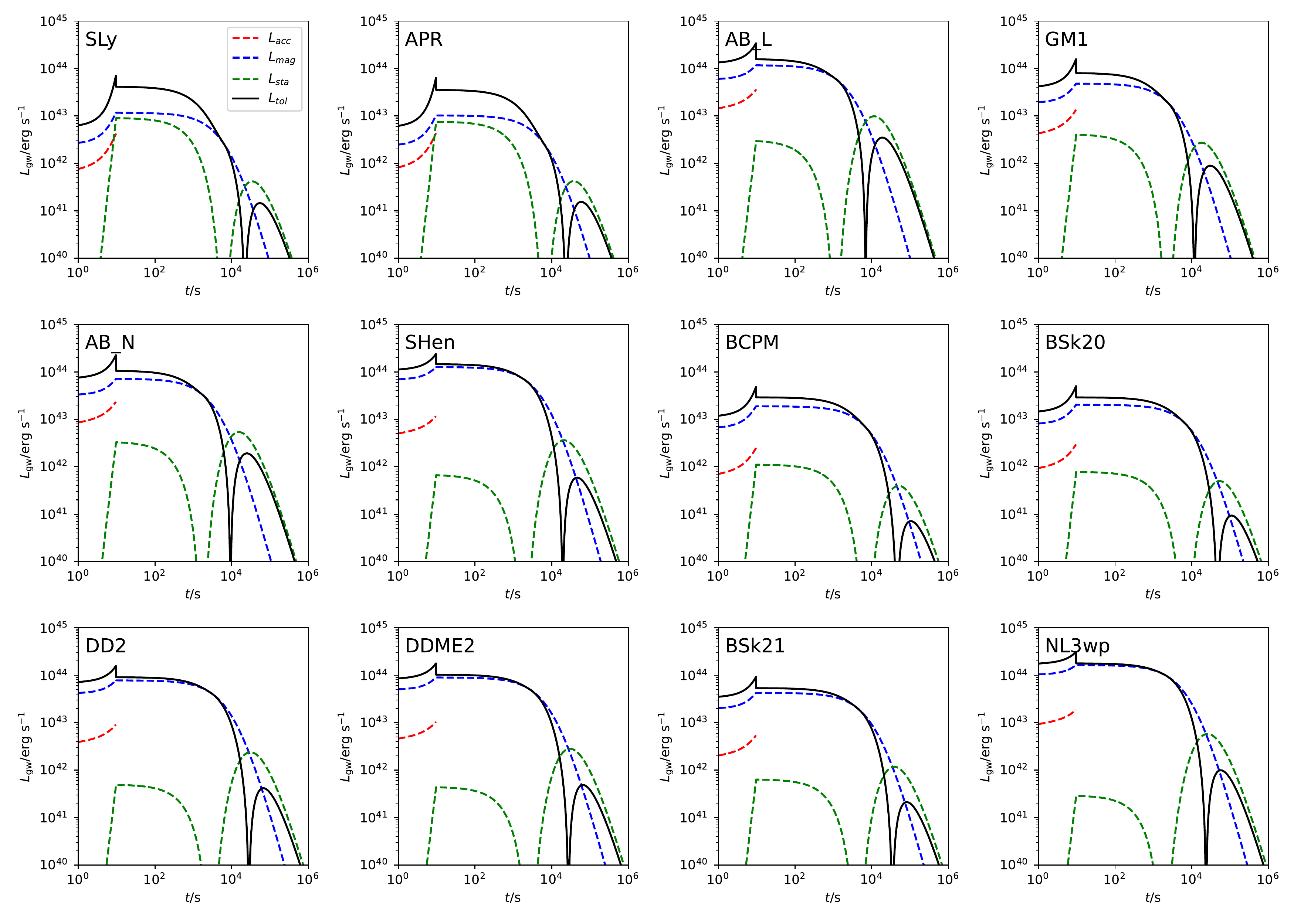}
		\caption{\label{fig:Lgw_EOS}Evolution of GW luminosity with different EOSs for given
			$B_0=5\times10^{14}\mathrm{G}$, $\Lambda=0.2$, and $\gamma=2.1$.}
	\end{figure*}

	As discussed above, we only adopt one magnetar EOS to do the calculations. Now, to discuss the
	dependence of the GW radiation on the EOS, we consider 12 EOSs that are reported in the
	literature~\cite{Lasky.et.al.2014, Ravi.et.al.2014,Ai.et.al.2018,Lan.et.al.2020,Lv.et.al.2021}.
	The relevant initial parameters we selected for different EOS are listed in
	Table~\ref{tab:table1}. The total mass of the neutron star will increase with the accretion, and the
	neutron star will collapse when the total mass of the neutron star is larger than $M_\mathrm{TOV}$.
	However, the initial mass of the neutron star remains uncertain. In order to test the long-lasting
	evolution of GW and avoid the neutron star collapse, we choose the initial mass of neutron star as
	$M_0=M_\mathrm{TOV}-0.2 M_{\odot}$ in our actual calculations. In fact, the masses observed by
	LIGO and Virgo in NS inspirals are quite high. The merger product is likely to be a supermassive or
	hypermassive NS which is supported by differential rotation. Further accretion coupled with
	spin-down would then lead it to collapse to a black hole. Therefore, our estimated of signal
	duration is optimistic and the signal in actual physical observations may be even shorter.
	Figure~\ref{fig:Lgw_EOS} shows the evolution of GW radiation with different EOSs for given 
	$B_0=5\times10^{14}$, $\Lambda=0.2$, and $\gamma=2.1$. We find that the GW radiation from magnetically 
	induced deformation, starquake deformation, and an accretion mountain are not very sensitive to 
	the EOSs we selected.
	\begin{table}[b]
		\caption{\label{tab:table1}Maximum mass of magnetar for different EOS and corresponding radius.}
		\begin{ruledtabular}
			\begin{tabular}{cccc}
				EOS             & $M_{\mathrm{TOV}}$ & $R$ & $I_0$ \\
				& $\left(M_{\odot}\right)$ & $(\mathrm{km})$ & $\left(10^{45} \mathrm{~g}
\mathrm{~cm}^{2}\right)$  \\\hline
				BCPM            & 1.98 & 9.94  & 2.86 \\
				SLy             & 2.05 & 9.99  & 1.91 \\
				BSk20           & 2.17 & 10.17 & 3.50 \\
				Shen            & 2.18 & 12.40 & 4.68 \\
				APR             & 2.20 & 10.00 & 2.13 \\
				BSk21           & 2.28 & 11.08 & 4.37 \\
				GM1             & 2.37 & 12.05 & 3.33 \\
				DD2             & 2.42 & 11.89 & 5.43 \\
				DDME2           & 2.48 & 12.09 & 5.85 \\
				AB-N            & 2.67 & 12.90 & 4.30 \\
				AB-L            & 2.71 & 13.70 & 4.70 \\
				NL3$\omega\rho$ & 2.75 & 12.99 & 7.89 \\
			\end{tabular}
		\end{ruledtabular}
	\end{table}
	
	\section{Detection Probability of a GW}
	One interesting question is that how strong of GW signal is from neutron star. In this
	section, we will present more details for calculation of the GW radiation.
	
\begin{figure}
		\includegraphics[width=3.5in]{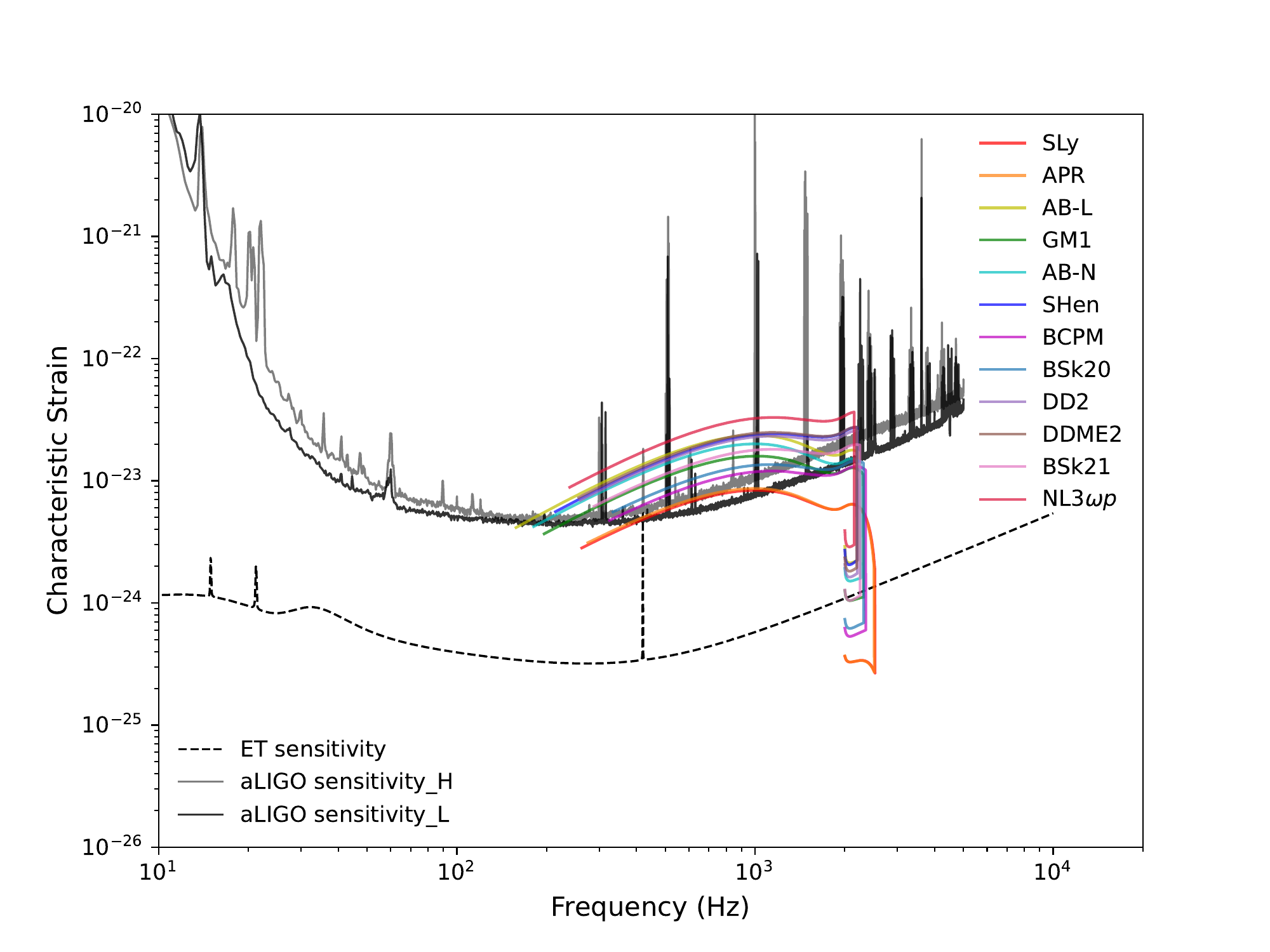}
		\includegraphics[width=3.5in]{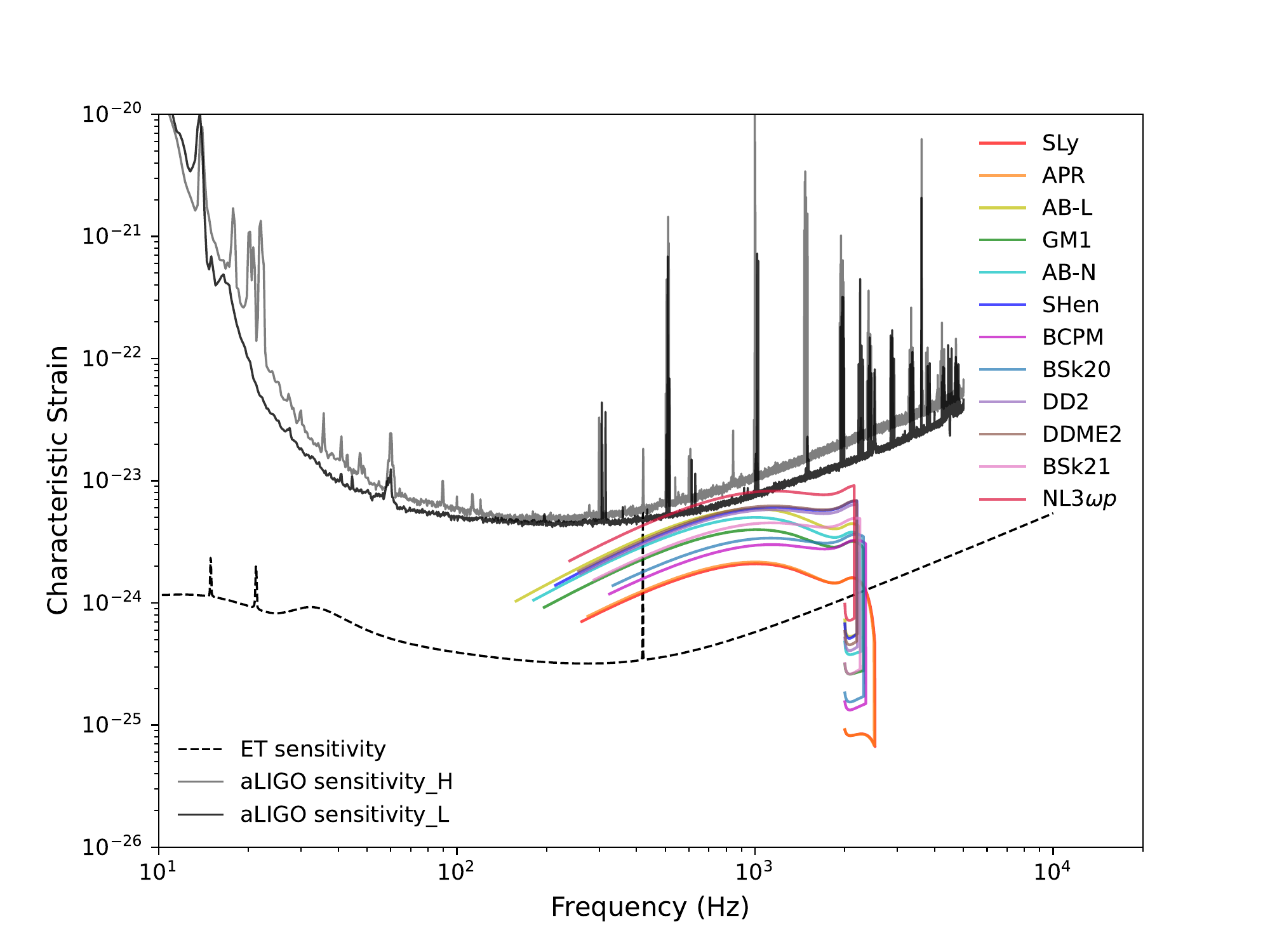}
		\caption{\label{fig:strian}
			Gravitational-wave strain evolution with frequency of magnetar for different EOSs at distance
			10 Mpc (top) and 40 Mpc (bottom). The black dotted line is the sensitivity limits for ET,
			and the black and gray solid lines are the
			sensitivity limits for aLIGO-Hanford and aLIGO-Livingston, respectively. The data of the
			noise curve are taken from Ref.~\cite{Lv.et.al.2021}.}
	\end{figure}

	The characteristic strain of GW from a rotating NS can be estimated
	as\cite{Corsi.et.al.2009,Hild.et.al.2011,Lasky.et.al.2016,Lv.et.al.2017}
	\begin{equation}
		h_{\mathrm{c}}=f h_0 \sqrt{\frac{d t}{d f}}
	\end{equation}
	and
	\begin{equation}
		\label{eq:h0}
		h_{0}=\frac{4 G I_0\epsilon}{D c^{4}}\Omega^{2}.
	\end{equation}
	$h_{0}$ is GW amplitude emitted by such an object at distance $D$, and $f=\Omega/\pi$ is the
	rotation frequency. Hence, combining with Eqs.~(\ref{eq:h0}) and (\ref{eq:domega}), the
	characteristic GW strain $h_{\mathrm{c}}$ can be rewritten as
	\begin{equation}
		h_{\mathrm{c}}=\frac{32 \pi^2 G I_0\epsilon}{ D c^{4}} \sqrt{\frac{\pi I_0} {N_{\mathrm{tol}}}}
f^{3}
	\end{equation}
	The GW strain is dependent on the distance, so we assume that the distance of magnetar is at 10 
	and 40 Mpc. By adopting the frequency range of GW from $f$=120 to 1000 Hz, one can estimate the
	maximum value of the strain $h_{\mathrm{c}}$ for different EOSs of NS at distance 10 and 40
	Mpc. In Fig.~\ref{fig:strian}, we plot the GW strain sensitivity for aLIGO and the ET. It is clear that the GW strain of magnetar with different EOSs is below the aLIGO
	noise curve at the distance of 40 Mpc, but it is expected to be detected by aLIGO at 10 Mpc. Moreover, we find
	that it is expected to be detected by the proposed ET in the future even for the distance of 40 Mpc.

	\section{Conclusion}
	A magnetar may survive for hundreds of seconds or longer after a binary neutron star merger or
	massive star core collapse. Weak GW radiation may be produced by the newborn magnetar due to its
	deformed structure. In this paper, we have investigated the evolution of the GW radiation of a
	magnetar by considering different deformations (e.g., magnetically induced deformation,
	starquake-induced ellipticity, and accretion column-induced deformation). The following interesting
	results are obtained.
	\begin{enumerate}[label=(\roman*)]
		\item For a given magnetar EOS, $B_0$, and $\Lambda$, the accretion torque and magnetic
		dipole radiation are dominant during the spin-up at early times and spin-down at later
		times, respectively, and the contribution of the GW radiation is small and can be ignored.
		\item For a given magnetar EOS and fixed $\Lambda=0.2$, the GW radiation signatures caused by
		an accretion mountain are stronger than that of starquakes and
		magnetically induced deformation for $B_0=10^{14}\mathrm{G}$. However, with the increased
		magnetic field strength, the GW radiation caused by magnetically induced deformation
		gradually becomes dominant.
		\item If the SLy EOS, $B_0=10^{15}$, and $\Lambda=0.2$ are fixed, the GW radiation from a
		magnetically induced deformation is dominant for $\gamma=2$. However, when we change
		$\gamma$ slightly from 2 to 2.1, the GW radiation of the starquake-induced deformation
		strengthens significantly and comparably with $L_{\rm mag}$ and $L_{\rm acc}$. A
		similar result also exists even when $\gamma$ increases to infinity.
		\item We selected 12 EOSs for given $B_0=5\times10^{14}$, $\Lambda=0.2$, and $\gamma=2.1$. We
		find that the GW radiation from magnetically induced deformation, starquake deformation,
		and an accretion mountain are not very sensitive to the different EOSs we selected.
		\item Finally, we calculate the GW strain with different EOSs, and find that it is
		difficult to be detected by aLIGO at 40 Mpc unless we move it closer to 10 Mpc. However, it
		is expected to be detected by ET in the future.
	\end{enumerate}
	
	In our calculations, we do not consider the evolution of the magnetic field when we calculate the
	GW radiation of magnetically induced deformation. There are two main reasons: One is that we do not
	know the details of the magnetic field evolution. The other is that the timescale of the
	magnetic field decay is long ($~10^4 \mathrm{yr}$)~\cite{Ho.et.al.2012}, and in fact is much longer
	than the lifetime of a newborn magnetar that we consider. Hence, we do not consider the evolution
	$\epsilon_\mathrm{mag}$ with the magnetic field.
	
	Although the GW radiation from a newborn magnetar has still not been detected by the current aLIGO
	and Virgo detectors, it plays a very important role in understanding the physics of neutron stars.
	The GW radiation is expected to be detected by the next generation of more sensitive GW detectors
	(e.g., ET), and a multimessenger detection will allow us to understand more details of the physics
	of neutron stars.

	\begin{acknowledgments}
		This work is supported by the National Natural Science Foundation of China (Grant No. 11922301 
		and No. 12133003), the Guangxi Science Foundation (Grant No. 2017GXNSFFA198008 and No. AD17129006), the
		Program of Bagui Young Scholars Program (L.H.J.), and special funding for Guangxi distinguished
		professors (Bagui Yingcai and Bagui Xuezhe).
	\end{acknowledgments}

	\nocite{*}
	
	\providecommand{\noopsort}[1]{}\providecommand{\singleletter}[1]{#1}%
	%

	
\end{document}